\documentclass[aip,pop,floats,reprint,superscriptaddress]{revtex4-1}
\usepackage{amssymb}
\usepackage{amsmath}
\usepackage{graphicx}
\graphicspath{{figures/}}
 \def\bc{\begin{center}}          \def\ec{\end{center}}
 \allowdisplaybreaks

\begin{document}
 \title{Dissipation of weakly nonlinear wakefields due to ion motion}
 \author{R.I.~Spitsyn}
 \affiliation{Budker Institute of Nuclear Physics SB RAS, 630090, Novosibirsk, Russia}
 \affiliation{Novosibirsk State University, 630090, Novosibirsk, Russia}
 \author{I.V.~Timofeev}
 \affiliation{Novosibirsk State University, 630090, Novosibirsk, Russia}
 \author{A.P.~Sosedkin}
 \author{K.V.~Lotov}
 \affiliation{Budker Institute of Nuclear Physics SB RAS, 630090, Novosibirsk, Russia}
 \affiliation{Novosibirsk State University, 630090, Novosibirsk, Russia}
 \date{\today}
 \begin{abstract}
In an initially uniform plasma, the lifetime of a weakly nonlinear plasma wave excited by a short driver is limited by the ion dynamics. The wakefield contains a slowly varying radial component, which results in a perturbation of the ion density profile and consequent destruction of the plasma wave. We suggest a novel method of characterizing the wave lifetime in numerical simulations quantitatively and study how the lifetime scales with the ion mass. We also discuss the implications of the limited lifetime on a recently proposed method of generating high-power terahertz radiation with counterpropagating wakefields driven by colliding laser pulses.
 \end{abstract}

 \maketitle

\section{Introduction}

The long-time evolution of regularly excited plasma waves is now of interest in the context of novel accelerating techniques, specifically laser- or particle beam-driven plasma wakefield acceleration.\cite{RAST9-19,RAST9-63,RAST9-85} The motion of plasma ions is one of the factors that determine the lifetime of the plasma wave. When the ion density is strongly perturbed due to the ion motion, the wave breaks, and the wave energy dissipates.\cite{PRL86-3332,PoP10-1124,PRL109-145005,PoP21-056705} This mechanism of wave termination becomes dominant for weakly nonlinear waves that would otherwise break at longer times.\cite{PoP21-056705,NIMA-829-3}

Our interest in the problem of wake behavior at ionic times arises from a novel theoretical idea of generating gigawatt narrow-band terahertz radiation by wakefields of counterpropagating laser pulses.\cite{Timofeev2017} In this scheme, colliding femtosecond laser pulses excite long-lived plasma oscillations capable of producing electromagnetic waves at the second harmonic of the plasma frequency $\omega_p$. In particle-in-cell simulations with immobile ions,\cite{Timofeev2017} the radiation continues for several tens of wave periods, as determined by depletion of the excited wakes. This time is sufficient for substantial perturbations of the initially uniform ion background to appear under the influence of plasma oscillations. In this paper, we study when and how the perturbations of the ion density destroy the plasma wave. We also discuss the effect of ion motion on the proposed radiation generation scheme.

Both qualitative and quantitative properties of the ion motion depend on plasma wave amplitude and structure, which in turn depend on laser beam parameters. We do not cover the whole diversity of possible regimes in this paper. Instead, we suggest a method of characterizing the wave lifetime and apply this method to studying a particular case of interest. We present a typical picture of wave dissipation at moderate wave amplitudes (about 20\% of the wavebreaking field) and study the dependence of the wave lifetime on the ion mass. The latter is important for experimental realization of the proposed radiation source, as it determines the acceptable time delay between the beams and the choice of the gas in which the laser beams collide. We focus on the evolution of a single wave excited by a single laser pulse. This is justified by the relatively low wave amplitude, for which the superposition principle holds, and enables us to use a fast quasistatic code for simulations.

The baseline case for the study corresponds to the proof-of-principle experiment under preparation in the Institute of Laser Physics SB RAS at Novosibirsk. In the experiment, two axially symmetric Gaussian laser pulses with the wavelength of 830 nm, duration of 39 fs, and total energy of 0.2 J will collide inside a supersonic gas jet. For the expected electron density of $2.5\times 10^{18}\ \mbox{cm}^{-3}$ after gas ionization, the radiation frequency is about 28 THz. To produce a terahertz pulse with a power of 30\,MW and total energy of 40\,$\mu$J, the first laser beam (170\,mJ) should be focused to the waist size of 40\,$\mu$m, while the waist size of the second laser beam (30\,mJ) should be as small as 15\,$\mu$m. The focal planes and propagation axes of both laser pulses will coincide. Because of the limited lifetime of the excited wakes, the laser pulses should arrive to the focus point within the limited time delay. The laser beams have the same peak intensity, but different transverse sizes, so the ponderomotive force is weaker in the wake of the first beam. This suggests that an earlier arrival of the first pulse to the focal plane would be preferable.

\begin{figure*}[htb]
\includegraphics[width=450bp]{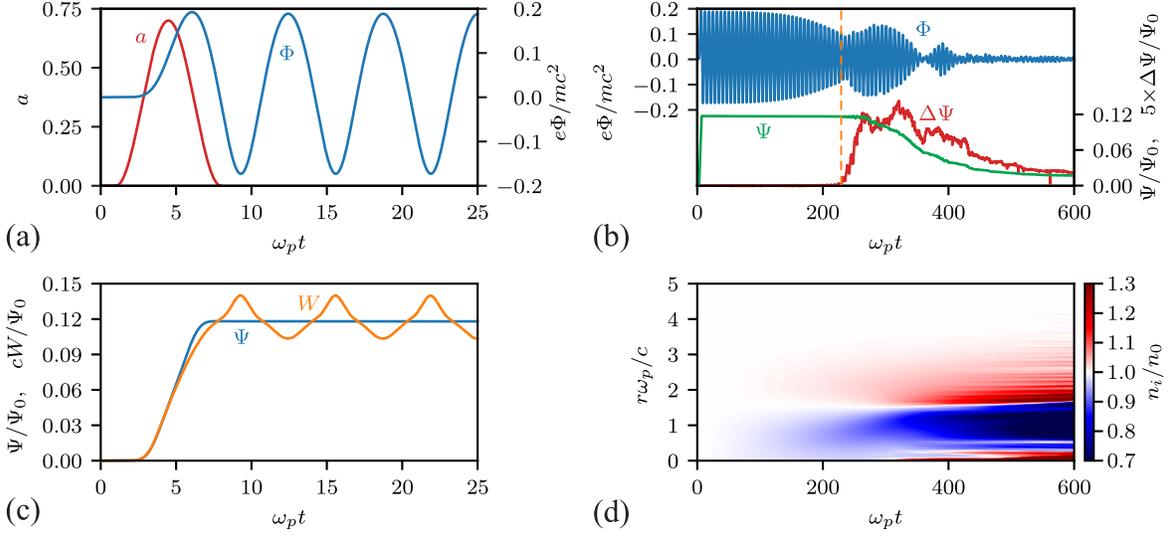}
\caption{(a) The wakefield potential $\Phi(t)$ (blue) and the laser strength parameter $a(t)$ (red) during the first few oscillation periods. (b) The wakefield potential $\Phi(t)$ (blue), the energy flux $\Psi(t)$ (green) and the difference $\Delta\Psi(t)$ enlarged 5 times (red) at a long time interval. The vertical dashed line indicates the onset of wave breaking. The energy flux unit $\Psi_0 = m_e^2 c^5/(4\pi e^2)$.  (c) The energy flux $\Psi(t)$ (blue) and the linear energy density $W(t)$ (orange) during the first few oscillation periods. (d) The radial distribution of the ion density versus time.}\label{fig1}
\end{figure*}

\section{The methods and the model}
\label{s2}

We simulate the temporal evolution of the plasma wave at the beam waist (at $z=0$) with quasistatic code LCODE.\cite{PRST-AB6-061301,NIMA-829-350} The laser pulse propagates in positive $z$ direction, is linearly polarized, has the frequency $\omega_0 = 25.44\,\omega_p$ and the envelope \\
\begin{equation}\label{e1}
	E_x= E_{0}  e^{-r^2/\sigma_s^2} \sin^2\left(\frac{\pi \xi}{c \tau}\right),  \quad 0 < t < \tau,
\end{equation}
where $\xi = z-ct$, and $c$ is the speed of light. The pulse duration $\tau \approx 7\omega_p^{-1}$ is optimal for wakefield excitation, and the maximum electric field amplitude $E_0$ corresponds to the laser strength parameter $a_0=eE_0/(m_e \omega_0 c)=0.7$, where $e$ is the elementary charge and $m_e$ is the electron mass. We consider two cases ($s=1,2$) that differ in waist sizes: $\sigma_1=5.87\,c/\omega_p$ and $\sigma_2=2.27\,c/\omega_p$. The laser intensity is high enough for complete ionization of hydrogen and helium and for fivefold barrier-suppression ionization of nitrogen. These are the gases we study in detail. The density of each gas corresponds to the fixed electron density $n_0=2.5\times 10^{18}\,\mbox{cm}^{-3}$. For illustrations, we use the narrower pulse ($s=2$) and helium plasma.

Figure~\ref{fig1} shows a typical picture of wave evolution. We characterize the wave amplitude with the wakefield potential $\Phi$. Its derivatives contain information about both accelerating and focusing strength of the wave,
\begin{equation}\label{e1a}
	\frac{\partial \Phi}{\partial \xi} = E_z, \qquad \frac{\partial \Phi}{\partial r} = E_r - B_{\phi},
\end{equation}
where $\vec{E}$ and $\vec{B}$ are electric and magnetic fields. The wave amplitude decreases neither steeply, nor exponentially [Fig.\,\ref{fig1}(b)], so the question arises of how to characterize the wave lifetime quantitatively. We suggest using the energy flux in the co-moving window.\cite{PRE69-046405} This flux $\Psi (\xi)$ can be defined under the conditions of the quasistatic approximation: the drive beam evolves slowly and the wakefield is almost stationary in the co-moving frame $(x,y,\xi)$. The flux is the difference of the linear energy density multiplied by $c$ and the usual energy flux in the laboratory frame. Its electromagnetic part is
\begin{equation}\label{e2}
    \Psi_{em} = \int dS \left( \frac{c}{8 \pi} (E^2 + B^2) - \frac{c}{4 \pi} \left[ \vec{E} \times \vec{B} \right]_z \right),
\end{equation}
where the integration is carried out over the transverse cross-section of the wake. The part associated with the kinetic energy arises from summing the contributions of individual particles that cross this section per unit time $\Delta t$,
\begin{equation}\label{e3}
    \Psi_p = \frac{1}{\Delta t} \sum_j (\gamma_j - 1) m_j c^2,
\end{equation}
where $m_j$ and $\gamma_j$ are the mass and relativistic factor of particles. Alternatively, we can consider plasma species as fluids, and calculate the fluid contribution to the kinetic energy
\begin{equation}\label{e4}
    \Psi_f = \int dS \sum_{s=i,e} \left( n_s m_s c^2 (\gamma_s - 1) (c - v_{sz}) \right),
\end{equation}
where $n_s$, $m_s$, $\vec{v}_s$, and $\gamma_s$ are the number density, particle mass, average velocity, and relativistic factor of a species. The total flux
\begin{equation}\label{e5}
    \Psi (\xi) = \Psi_{em} + \Psi_p
\end{equation}
is constant in the absence of energy sources and sinks. For $z=0$ being fixed, it becomes a function of time $t$. It increases as the laser pulse drives the wave [Fig.\,\ref{fig1}(c)] and, in the absence of a witness bunch, decreases only when some high-energy particles reach the boundary of the simulation window soon after the wave breaks [Fig.\,\ref{fig1}(b)]. For this reason, the energy flux is a more convenient measure of the wave energy than the usual linear energy density
\begin{equation}\label{e6}
    W = \int \frac{E^2 + B^2}{8 \pi} dS + \frac{1}{\Delta z} \sum_j (\gamma_j - 1) m_j c^2,
\end{equation}
where the summation is over all plasma particles in a narrow layer of the thickness $\Delta z$. The linear energy density is not constant because the plasma wave causes longitudinal energy flows both into and out of the considered layer [Fig.\,\ref{fig1}(c)].

Fluid $\Psi_f$ and particle $\Psi_p$ contributions to the energy flux coincide only in a cold plasma. As the wave breaks, they start to differ because of multiple flow:
\begin{equation}\label{e7}
    \Delta \Psi = \Psi_p - \Psi_f \neq 0.
\end{equation}
This difference abruptly rises at some point [Fig.\,\ref{fig1}(b)], indicating the onset of wave breaking. Shortly after the appearance of nonzero $\Delta \Psi$, fast plasma electrons exit the simulation window radially, and we observe a bend of the $\Psi (t)$ curve [Fig.\,\ref{fig1}(b)]; the delay between these two events depends on the simulation window radius. After a while the wave disappears. We define the wave lifetime as the moment at which $\Delta \Psi$ exceeds 1\% of the maximum value of $\Psi (t)$. The proposed method of wavebreaking identification is more practical than detection of multiple flows (particle trajectories' intersection), as the latter may be caused by initial plasma temperature or numerical plasma heating.

\begin{figure}[b]
\includegraphics[width=225bp]{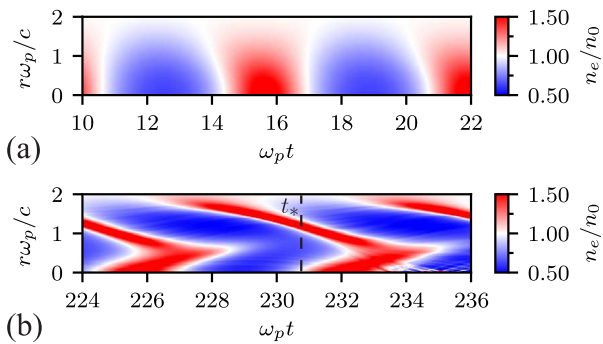}
\caption{Radial distributions of the electron density $n_e$ shortly after the driver passage (a) and near the wavebreaking moment (b).}\label{fig2}
\end{figure}

\begin{figure}[t]
\includegraphics[width=237bp]{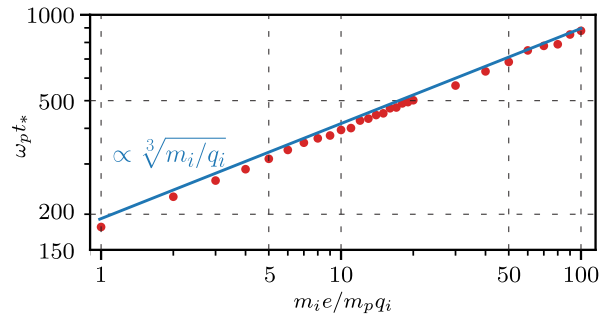}
\caption{The wavebreaking time $t_*$ versus ion mass-to-charge ratio $m_i/q_i$ (in units of proton mass-to-charge ratio $m_p/e$) obtained from theory (solid line) and simulations (points).}\label{fig3}
\end{figure}

\section{Simulation results}\label{s3}

Once we defined the wave lifetime, we can study its scalings and the process of wakefield dissipation. At the considered amplitudes, the wave breaks because of distortion of the ion background\cite{PRL86-3332,PoP10-1124} [Fig.\,\ref{fig1}(d)]. The perturbation of the ion density results in different electron oscillation frequencies at different radial positions and, therefore, distortion of phase fronts (Fig.\,\ref{fig2}). At the moment $t_*$ of wavebreaking, the relative phase shift between the areas of the lowest ion density (at some $r\lesssim c/\omega_p$) and unperturbed density (at large $r$) approaches $2 \pi$ [Fig.\,\ref{fig2}(b)]. Then a group of particles falls out of coherent oscillations, crosses the axis, and continues moving ballistically to the outer boundaries of the simulation window. The wave amplitude does not decrease monotonically after the wavebreaking. On the contrary, the longitudinal field grows for a short period, and only then decays [Fig.\,\ref{fig1}(b)]. Simulations of the wave driven by a self-modulating proton beam also reveal a similar behavior (Fig.\,12 in Ref.~\onlinecite{NIMA-829-3}), which is explained by the wave energy concentrating near the axis due to ion density gradients.\cite{Minakov}

The theory of wave breaking in the presence of ion motion was developed in Refs.~\onlinecite{PRL86-3332,PoP10-1124}, but was not benchmarked against simulations for realistic ion masses. Now we can extend this comparison to a wide interval of ion mass-to-charge ratios (Fig.\,\ref{fig3}). The theoretical prediction in Fig.\,\ref{fig3} is the formula (33) from Ref.\,\onlinecite{PoP10-1124} adapted for our laser pulse profile. In simulations, the wave breaks one or two periods earlier than the theory predicts, which is good precision considering that in simulations $t_*$ can take only discrete phase-bound values.

\begin{figure}[tb]
\includegraphics[width=219bp]{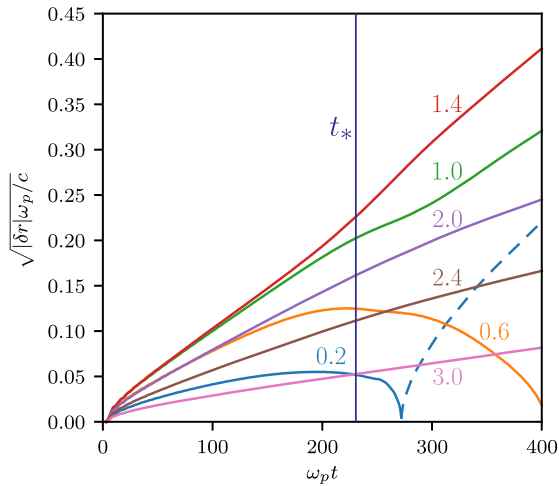}
\caption{Time dependencies of ion displacement. For better visibility of the linear acceleration, we show the square root of ion displacement $\delta r$. Numbers next to the curves correspond to initial radii of ions in units of $c/\omega_p$. The dashed line denotes $\delta r < 0$.}\label{fig4}
\end{figure}

The observed $m_i^{1/3}$ scaling of the wavebreaking time has a simple explanation. The radial force acting on the ions does not depend on the ion mass. It is often called ponderomotive force, though, strictly speaking, it is the force of charge separation field, which is equal to the ponderomotive force exerted on electrons. This force linearly accelerates the ions in positive $r$ direction and changes their positions by $\delta r \propto t^2/m_i$ (Fig.\,\ref{fig4}). Despite the fact that the ions experience different acceleration at different radii, the change of the ion density $\delta n_i$ and the elongation of the wave period $\delta \lambda_p$ has the same scaling $|\delta n_i| \propto \delta \lambda_p \propto t^2/m_i$ as long as the perturbations remain small. The wave breaks when the cumulative phase advance becomes of the order of $2\pi$, or if the integral
\begin{equation}\label{e9}
    \int_0^{t_*} \delta \lambda_p (t) \, dt \propto \frac{t_*^3}{m_i}
\end{equation}
reaches some threshold value, whence
\begin{equation}\label{e10}
    t_* \propto m_i^{1/3}.
\end{equation}

The time dependencies of ion displacement at different radii (Fig.\,\ref{fig4}) not only prove the linearity of acceleration, but also demonstrate other important features of ion motion. We see that the initial push given to ions by the driver is negligibly small compared to the effect of the wakefield, as predicted in Ref.\,\onlinecite{PoP10-1124}. After the wave breaks, the field structure changes, and near-axis ions reverse the direction of motion and proceed to form a density increase near the axis [Fig.\,\ref{fig1}(d)].

\begin{table}[b]
\centering
\caption{Simulated wave lifetimes $t_{s*}$ for two considered laser pulses ($s=1,2$) in various plasmas; maximum delays $t_{D} = t_{1*}-t_{2*}$ between the pulses.}\label{t1}
\begin{tabular}{ lccc }
 \hline\hline
 Plasma \ & \ $\omega_pt_{1*}$ \ & \ $\omega_pt_{2*}$ \ & \ $\omega_p t_D$ \\
 \hline
 $\text{H}^+$ & 383.6 & 180.3 & 203.3 \\
 $\text{He}^{2+}$ & 487.4 & 230.3 & 248.1 \\
 $\text{N}^{5+}$ & 528.2 & 261.2 & 267.0 \\
 \hline
 \hline
\end{tabular}
\end{table}

\section{Synchronization requirements}

Let us now determine what time delay between arrivals of laser pulses to the focus point is tolerable in the proposed experiment on terahertz generation. Earlier particle-in-cell simulations with immobile ions\cite{Timofeev2017} have shown that the duration of intense radiation is limited by about 100\,$\omega_p^{-1}$. This time is shorter than the limit imposed by the wavebreaking even for the narrower pulse in hydrogen (Table~\ref{t1}). The wider wake lives longer, so the first laser pulse may arrive to the focus point earlier than the second one. The maximum time delay $t_{D}$ between the two pulses under the condition of maximum radiation duration is achieved if both waves break at the same time. For the hydrogen, the maximum delay is 203.3\,$\omega_p^{-1}$ (2.3\,ps). For heavier atoms, it is slightly longer. The most appropriate candidates for our scheme are helium and nitrogen. The considered laser pulses are able to produce a wide enough plasma channel with a uniform electron density corresponding to fully ionized helium or fivefold ionized nitrogen. In heavier gases such as argon or xenon, the electron density is expected to be nonuniform across the plasma channel at the ionization stage, which can result in faster destruction of the excited wakes. Consequently, the tolerable time delay between the arrivals of two laser pulses to the focus point should be less than or of the order of 1\,ps.

\section{Conclusion}

We proposed a novel way to determine the lifetime of a plasma wake excited by a short driver. In our method, the moment of wave breaking is associated with appearance of fast particles and quick modification of the energy flux in the co-moving frame. By analyzing the temporal evolution of wakefields excited by femtosecond laser pulses, we show that the radial ion dynamics inside a plasma wake can be crucial for experimental implementation of a recently proposed scheme of terahertz generation, in which the electromagnetic emission is produced by counterpropagating plasma wakes of differing radial profiles. In particular, arrival of laser pulses to the focal point should be synchronized within 1\,ps.

\section{acknowledgments}
This work is supported by RFBR grant 18-42-540010. Simulations are carried out using computational resources of Novosibirsk State University and Siberian Supercomputer Center SB RAS.

\end{document}